\documentstyle[preprint,aps]{revtex}
\begin{document}

\newcommand{\btu}{\bigtriangledown}
\newcommand{\bge}{\begin{equation}}
\newcommand{\ege}{\end{equation}}
\newcommand{\bga}{\begin{eqnarray}}
\newcommand{\ega}{\end{eqnarray}}
\newcommand{\nnu}{\nonumber}

\draft
\preprint{\vbox{\hbox{IP/BBSR/95-59}\hbox{hep-th/9506213}\hbox{June,
1995}}}
\title {  SYMMETRIES OF FOUR-DIMENSIONAL STRING EFFECTIVE ACTION
WITH COSMOLOGICAL CONSTANT}
\author{ Jnanadeva Maharana \cite{mail1} and Harvendra Singh \cite{mail2} }
\address{Institute of Physics, Bhubaneswar-751 005, INDIA.}
\maketitle
\begin{abstract}

Classical solutions for  a four-dimensional Minkowskian string effective
action and an Euclidean one  with cosmological constant term are derived. The
former corresponds to electrovac solutions whereas the later solutions are
identified as gravitational instanton solutions for Fubini-Study metric.
The symmetries of the effective actions are identified and new classical
solutions are generated by implementing appropriate noncompact
transformations. The S-duality transformations on the equations
of motion are discussed and it is found that they are S-duality
noninvariant due to the presence of cosmological constant term.
\end{abstract}


\narrowtext

\newpage

\par The target space symmetries of string theory such as T-duality,
$O(d,d)$ transformations and S-duality
have attracted considerable attention in the recent
past. Let us recall some salient aspects of T-duality: if we
consider a compactified string on a circle of radius $R$, the
spectrum of the string remains invarient under the
transformation $R\to {1\over R}$. In a more general scenario,
when we envisage evolution of a string in the background of its
massless excitations and compactify $d$ of its coordinates on a
torus, a richer symmetry structure emerges if all these
backgrounds are independent of those compactified coordinates.
The reduced string effective action has an invariance under global symmetry
transformations, $O(d,d)$. The T-duality belongs to discrete
subgroup of $O(d,d)$ transformations.
The target space duality ( T-duality ) is the generalisation of
$R\to{1\over R}$ transformation.  The metric that
appears in the reduced string effective action is the
$\sigma$-model metric.  Under  T-duality \cite{ky}
and $O(d,d)$ transformations \cite{gr},
the space-time metric and the ( shifted ) dilaton remain
invariant. It is well known that under T-duality and $O(d,d)$
transformations, one can generate new background configurations
starting from a known solution of the string effective action.
On the other hand, study of S-duality \cite{stw,ss} in string
theory holds the prospect of revealing the nonpertervative
features of string theory. It is possible to relate strong and
weak coupling regimes of string theory under this transformation.
   In four dimensions, the axion and dilaton can be combined to define
a complex scalar field which transforms under $SL(2,{\bf Z})$
group while leaving equations of motion unaltered. However,
it is more convenient to go over to the Einstein metric while discussing
S-duality transformations and their consequences.

   \par The purpose of this note is to derive classical
   solutions of four-dimensional
string effective actions. Then we generate new solutions through $O(d,d)$
transformations. First we consider a Minkowskian action where solution was
derived for $N=2$ supergravity action which was designated as
``electrovac'' solution \cite{fg}. However, the dilaton field
appears in the massless multiplet of the string theory and solutions of the
string effective action are required to  satisfy  equations of motion
associated with
each massless mode of the string. Therefore, the presence of the dilaton field
in
the action imposes additional constraints on the background fields. We shall
present solutions in the presence of constant dilaton background. However,
this effective action has an invariance under noncompact
symmetries. Thus we shall obtain  new background fields from the known
classical solution by suitably implementing noncompact symmetry
transformations.
We recall that  T-duality and $O(d,d)$ transformations have been applied
to obtain new solutions in the context of string cosmological
solutions\cite{gmv1,sen2}, black holes
\cite{ag,ek,kks,sen3,sen4,bmq,hs} and topology changing processes
\cite{km,ck}.

\par The second problem we consider is the Euclidean effective action in
four-dimension in the presence of Abelian gauge fields. This action admits
classical solutions such that Weyl curvature tensor and the gauge field
strength
satisfy self-duality condition and the solution has the interpretation
of gravitational instanton \cite{gh}.
Furthermore, we show that new metric and gauge field configurations can be
generated by  a global noncompact transformation.
However, the new backgrounds are such that they do not satisfy self-duality
condition.

We would like to mention that the two effective actions that are
considered in obtaining the classical solutions have
cosmological constant terms. We shall see later that the
equations of motion do not remain invariant under S-duality
transformations due to the presence of cosmological constant term
( we go over to Einstein frame when we examine this issue ).

Let us recall the most salient aspects of dimensional reduction of the string
effective action. We write the low energy heterotic string effective action
in D dimensions with Abelian gauge fields following ref.\cite{gr},

\bga
S& =& \int d^DX^M \sqrt{\det {\hat G}_{M N}}\, e^{-{\hat\Phi}}\nnu\\
&&\left( R_{\hat G}+{\hat G}^{M N}
\partial_M{\hat\Phi}\partial_N{\hat\Phi} -{1\over4}{\hat F}^{I\,M N}{{\hat
F}^I}_{M N}
-{1\over12}{\hat H}_{M N P}{\hat H}^{M N P} -2\Lambda \right)
\label{11}
\ega
where
\bga
{\hat H}_{M N P} &=& \partial_M{\hat B}_{N P} + {\rm cyclic\,\,
permutations},\nnu\\
{\hat F}^I_{M N} &=& \partial_M {\hat A}^I_N - \partial_N{\hat A}^I_M \nnu\\
M\,,\,N &=& 1,\cdots,D; \,\,\,\,\, I=1,\cdots,n \,.
\label{12}
\ega
Here ${\hat G}_{M N}$, ${\hat\Phi}$, ${\hat A}^I_M$ and ${\hat B}_{M N}$ denote
the graviton,
dilaton, n-component Abelian vector field and antisymmetric tensor fields,
respectively.
$R_{\hat G}$ denotes D dimensional scalar curvature and $\Lambda$ is the
deficit in
central charge which plays the role of cosmological constant. In dimensional
reduction scheme, for backgrounds independent of d coordinates ( say,
$X^\alpha,\,1\le \alpha
\le d$ ), with toroidal compactification on $T^d$, the action (\ref{11}) can be
rewritten as

\bga
S &=& \int d^{D-d}X^\mu\sqrt{\det g_{\mu\nu}}\,\,e^{-\Phi} \nnu\\
&&\left( R_g + g^{\mu\nu}\partial_\mu\Phi\partial_\nu\Phi + {1\over8} Tr
\partial_\mu M^{-1}
\partial^\mu M -{1\over4} {\cal F}^i_{\mu\nu} ( M^{-1} )_{i\,j}{\cal
F}^{j\,\mu\nu}
-{1\over12} H_{\mu\nu\lambda}\,H^{\mu\nu\lambda} - 2\Lambda \right)
\label{13}
\ega
where
\bga
   {\hat G}_{M N}&=&\pmatrix{ g_{\mu\nu} + A^{(1)}_{\mu\,\alpha}
   A^{(1)\,\alpha}_\nu & A^{(1)}_{\mu\,\beta}\cr
   A^{(1)}_{\nu\,\alpha} & G_{\alpha\beta}}\nnu\\ \nnu\\
   \Phi &=& {\hat \Phi} - {1\over2}\ln\det G_{\alpha\beta} \nnu\\ \nnu\\
M &=& {\cal L} M^{-1}{\cal L} = \pmatrix{ G^{-1} & -G^{-1} C & -G^{-1} A^T \cr
-C^T G^{-1} & G+C^T G^{-1} C + A^T A & C^T G^{-1}A^T + A^T \cr
-A G^{-1} & A G^{-1}C +A & 1+ A G^{-1} A^T}\nnu\\ \nnu\\
C_{\alpha\beta}&=& {1\over2} A^I_\alpha A^I_\beta + B_{\alpha\beta},
\label{14}
\ega
with the space-time dependent background fields $( G_{\alpha\beta}\,,
\,\, A^I_\alpha
\equiv{\hat A}^I_\alpha\,,\,\,B_{\alpha\beta}\equiv {\hat B}_{\alpha\beta} )$
defining a generic point in moduli-space in the toroidal
compactification of the heterotic string theory. Note that moduli M
satisfies the condition $ M {\cal L} M {\cal L} =1$,
 where ${\cal L}$ is the $O(d,d+n)$ metric,

\begin{eqnarray}
{\cal L} = &&\pmatrix{ 0 & I_d & 0 \cr I_d & 0 & 0\cr 0 & 0 &I_n}\nnu\\
\Omega^T {\cal L}&& \,\Omega = {\cal L}\,.
\label{25}
\end{eqnarray}
Here $I_d$ is d-dimensional identity matrix and $\Omega$ is an element of
the group $O(d,d+n)$ . The definitions of other two fields in (\ref{13}) are

\bga
H_{\mu\nu\lambda}&=& \partial_\mu B_{\nu\lambda} -{1\over2}{\cal A}^i {\cal
L}_{i\,j}
{\cal F}^j_{\nu\lambda} + {\rm cyclic\,\, permutations}\nnu\\
{\cal F}^i_{\mu\nu}&=& \partial_\mu {\cal A}^i_\mu - \partial_\nu{\cal A}^i_\mu
\label{15}
\ega
where i, j are $O(d,d)$ matrix indices. ${\cal A}^i_\mu = (
\,A^{(1)\alpha}_\mu, \,A^{(2)}_{\mu\alpha}, \,
A^{(3)\,I}_\mu \,)$ is a $(2d +n )$ component vector field with the following
definition
of its components,
\bga
A^{(1)}_{\mu\alpha}&=& {\hat G}_{\mu\alpha}\nnu\\
A^{(2)}_{\mu\alpha}&=& {\hat B}_{\mu\alpha} + {\hat B}_{\alpha\beta}
A^{(1)\beta}_\mu
+{1\over2} {\hat A}^I_\alpha \, A^{(3)\,I}_\mu \nnu\\
A^{(3)\,I}_\mu&=& {\hat A}^I_\mu - {\hat A}^I_\alpha\, A^{(1)\,\alpha}_\mu\,.
\label{16}
\ega
   Now it is straightforward to check that the action (\ref{13}) is
   manifestally invariant under global $O(d,d+n)$ transformations
 \bga
 M&\to& \Omega \, M \, \Omega^T\nnu\\
 \Phi&\to& \Phi,\,\,\, g_{\mu\nu}\to g_{\mu\nu},\,\,\,,B_{\mu\nu}\to
B_{\mu\nu},\nnu\\
 {\cal A}^i_\mu&\to& \Omega^i_j \,{\cal A}^j_\mu
 \label{17}
 \ega
 where $\Omega$ is an $O(d,d+n)$ matrix such that
$\Omega {\cal L} \Omega^T={\cal L}$.
Note that ${\cal A}^i_\mu$ transforms as a vector multiplet under $O(d,d+n)$
transformations.
We mention in passing that $M\to M^{-1}$ is the analog of
$R\to{1\over R}$ duality for the general case.
In general, an $O(d)\times O(d+n)$ transformation, which is a
subgroup of $O(d,d+n)$ generates new
solutions which cannot be obtained from the old ones
through general coordinate transformations or
   gauge deformations. There are specific examples where an $O(d)\times O(d)$
transformations can "boost away" curvature singularities and in fact can
turn flat string backgrounds into nonflat ones and vice versa. It has been
shown explicitely in reference \cite{gmv2} that Nappi-Wtten
backgrounds in four-dimension can be obtained from
$O(2,2)$ boosting of the direct product of a pair of
two-dimensional models.

\par We consider the following four-dimensional string effective action

\bga
   S&=&\int d^4X^M \sqrt{\det {\hat G}_{M N}}\,
   e^{-{\hat\Phi}}\left(
   R_{\hat G}+{\hat G}^{M N}
\partial_M{\hat\Phi}\partial_N{\hat\Phi} -{1\over4}{\hat F}^{M N}{{\hat F}}_{M
N}
+2 g_B^2 \right)
\label{21}
\ega
The equations of motion are satisfied for following background field
configurations,

\begin{eqnarray}
{\hat ds}^2 &=& dx^2 + dy^2 + d\xi^2 - \cosh^2[\sqrt{2} g_B\xi] \,
dt^2\nonumber\\
{\hat A}_M &=&\pmatrix{ 0 & 0 & 0 & \sqrt{2}  \sinh[ \sqrt{2} g_B \xi]}\nnu\\
{\hat\Phi} &=& {\rm constant}\,.
\label{22}
\end{eqnarray}
This effective action is similar to the $N=2$ supergravity action considered by
Freedman and Gibbons\cite{fg} with appropriate choice of backgrounds. The
solution given by
(\ref{22}) are analog of the ``electrovac'' solutions of ref.\cite{fg} .
Notice, however,
that massless dilaton appears in the action (\ref{21}). Now the
dilaton equation of motion is required to be satisfied in
addition to Einstein and Maxwell field equation. It is
worthwhile to point out that this equation, even for constant
dilaton background, imposes constraints on the backgroud field
equations. The topology of the spacetime is $ R^2
\times AdS_2$ with the curvature scalar $R = -4 g_B^2$ .
The Maxwell fild strength is covariantly
constant, $i.e.$ $ \btu_M {\hat F}^{M N}=0$, and is  of purely
electric-type
\bga
{\hat F}_{\xi\, 0} = 2 \,g_B \,\cosh[\sqrt{2}\, g_B \,\xi],
\,\,\,\,\,\>{\hat F}_{i j}=0,\,\,\,\,\,i,j = x,y,\xi\,\,\,,
\nnu
\ega
and the field strength satisfies the constraint $ {\hat F}_{M N}
{\hat F}^{M N}= -8 g_{B}^2 \,$.

Note that backgrounds in (\ref{22}) are independent of the coordinates
x, y and t. Therefore action (\ref{21}) whose contents are
graviton, dilaton and an Abelian gauge field is invariant under
 $O(3,4)$ symmetry transformations.
Thus one can perform $O(3,4)$ trasformations on this
background  to obtain new classical
solutions which can generate a nontrivial antisymmetric tensor field
${\bar B}_{M N}$. For sake of definiteness, we shall choose a
specific transformations restricted to $t-y$ plane only \cite{jm}.
This amounts to restricting ourselves to $O(2,3)$ which is a
subgroup of $O(3,4)$.
Let us first rewrite action (\ref{21}) isolating y- and t-isometries ,
\bga
S&=&\int dy dt \int dx d\xi \sqrt{\det g_{\mu\nu}} \,e^{-\Phi}\nnu\\&& \left(
R_{g} +
\partial_\mu \Phi \partial^\mu \Phi +{1\over8} Tr(\partial_\mu
M^{-1}\partial^\mu M) -{1\over4} {\cal F}_{\mu\nu}^{(i)}
(M^{-1})_{i j} {\cal F}^{(j) \mu\nu} + 2 g_{B}^2 \right)
\label{26}
\ega
where,
\bga
g_{\mu\nu}&=& \delta_{\mu\nu} ;\>\>\>\>\>\>\>\>\mu,\nu=x,\xi
\nnu\\
\Phi &=& {\hat\Phi} - {1\over2} \ln \det G_{\alpha\beta} \nnu\\ \nnu\\
G_{\alpha\beta}&=&\pmatrix{ -\cosh^2[\sqrt{2} g_B \xi] & 0 \cr
0 & 1};\>\>\>\>\>\>\>\>\alpha,\beta= t,\,y \nnu\\ \nnu\\
A_\alpha &=& ( \,\,\,\sqrt{2} \sinh[\sqrt{2} g_B
\xi]\>\>\>,\>\>\>0 \,\,\,) \nnu\\
B_{\alpha\beta}&=&0 \nnu\\
{\cal F}^{(i)}_{\mu\nu}&=& 0 \,.
\label{27}
\ega
The moduli matrix $M$ can be constructed using the definition given in
eq. (\ref{14}). To obtain new background we choose $\Omega$ to be of the
following form
\bge
\Omega=\Omega_1 \,\times \,\Omega_2
\label{28}
\ege
where
\bga
\Omega_1 &=&{1\over2}\pmatrix{ c+1 & -s & 1-c & -s & 0 \cr -s & c+1 & s &
c-1 & 0 \cr 1-c & s & 1+c & s & 0 \cr  -s &
c-1 & s & c+1 & 0 \cr 0 & 0 & 0 & 0 & 2 }\nnu\\ \nnu\\ \nnu\\
\Omega_2 &=&{1\over2}\pmatrix{ {\hat c}+1 & 0 & 1-{\hat c} & 0 &
-\sqrt{2} {\hat s} \cr
0 & 1 & 0 & 0 & 0 \cr 1-{\hat c} & 0 & {\hat c}+1 & 0 & \sqrt{2}{\hat s} \cr 0
& 0 &
0 & 1 & 0 \cr - \sqrt{2} {\hat s} & 0 & \sqrt{2}{\hat s} & 0 & 2{\hat c}}
\label{29}
\ega
and $c=\cosh\theta,s=\sinh\theta,{\hat c}=\cosh\gamma$, and
${\hat s}=\sinh\gamma$ with $0< \,\theta < \infty$, $0<
\gamma\,< \infty$ being the boost parameters. Notice that our
rotations involve the $t-y$ plane and acts on $G_{\alpha\beta}$ and the
gauge field $A_\alpha$. These two noncompact rotations, generically,
called ``boosts'',
appear because the $t$-coordinate is involved. The ``boosts"
$\Omega_1$ and $\Omega_2$ form the elements of the
group $O(1,1)\times O(1,2)$ which is a subgroup
of $O(2,3)$. The backgrounds thus generated
satisfy the equations of motion and are not connected to the
original background configuration by general coordinate transformations
and/or gauge transformations.

We note that the metric and the gauge potential have the
following form after $O(1,1)\times O(1,2)$ transformations
summarised in (8),
\bga
{\bar {ds}}^2&=& dx^2 + d\xi^2 - {1 + c^2 a(\xi)^2 \over (1+ c\,{\hat s}
a(\xi))^2}dt^2 +{2 s a(\xi)(c a(\xi) - {\hat s})\over (1+ c\,{\hat s}
a(\xi))^2} dy \, dt\nnu\\& &+ {1 + 2 c\,{\hat s} a(\xi) + ({\hat c}^2 -
c^2)a(\xi)^2 \over (1+ c\,{\hat s}
a(\xi))^2} dy^2 ,\nnu\\ \nnu\\
{\bar A}_M &=& (\,\,\,0,\,\,\,0,\,\,\, {\sqrt{2} c\,{\hat c}
a(\xi) \over 1+ c\,{\hat s}
a(\xi)}\,\,,\,\,-{\sqrt{2} s\,{\hat c} a(\xi) \over 1+ c\,{\hat s}
a(\xi)}\,\,\,) \,.
\label{30.1}
\ega
where $a(\xi)=\sinh[\sqrt{2} g_B \xi]\,$. Furthermore, the
antisymmetric tensor field strength is nonzero after the $O(1,1)\times
O(1,2)$ transformations ( although, the original background had
vanishing ${\hat H}_{M N P}\,$),

\bge
{\bar B_{t\,y}}= {s\,{\hat s} a(\xi) \over 1+ c\,{\hat s}
a(\xi)}\,.
\label{30.2}
\ege
Moreover, we find that the transformed dilaton depends on coordinate $\xi $
nontrivially and is given by
\bge
{\bar  \Phi}= {\hat \Phi} - \ln ( 1 + c\,{\hat s} a(\xi))\,.
\label{30.3}
\ege
The new Maxwell field strength, ${\bar F}_{M N}$, has both
electric and magnetic components and ${\bar F_{M N}}\,{\bar F^{M
N}} = -8 g_B^2 { {\hat c}^2 \over (1 + c \,{\hat s}\, a(\xi))^2}$.

The curvature scalar is
\bge
{\bar R} = -g_B^2 { {\cal R}(\xi) \over ( 1 + c\,{\hat s}\,a(\xi) )^6}
\label{30s}
\ege
where the numerator
\bga
{\cal R}(\xi)&=&  4 + {\hat s}^2(1 + 7 c^2) + (12 + 4 {\hat s}^2 (1+7\, c^2)
)\,c\,{\hat s} \,a(\xi) \nnu\\
&&+ ( 8 + 6 {\hat s}^2 ( 1+ 7 c^2))\,c^2\,{\hat s}^2\,a(\xi)^2 -
(8 - 4 {\hat s}^2 ( 1+7 c^2))
\,c^3 \,{\hat s}^3 \, a(\xi)^3 \nnu\\
&&-(12 - {\hat s}^2(1+7 c^2))\, c^4 \,{\hat s}^4\,a(\xi)^4
- 4 c^5\, {\hat s}^5 \,a(\xi)^5 \,.\nnu
\ega
The new background is singular when $( 1+ c\,{\hat s}\,a(\xi) )
= 0\,$ while the original background had a constant curvature, $-4 g_B^2\,$.

Now we turn to discuss S-duality transformations.
The dilaton and axion $\,\chi\,$ ( dual to the antisymmetric tensor field
in four dimensions ) can be combined to define a complex scalar
field, $\Psi = \chi + i \,\eta\, $ , where $ \eta = e^{-\Phi}\,$.
The S-duality transformations \cite{ss} correspond to
\bga
&&\Psi \to {a\,\Psi + b \over c\,\Psi +
d}\,,\,\,\,\,\,\,a,b,\cdots\epsilon \,\,{\bf Z}\nnu\\
&&{\hat F}_{M N} \to c \,\eta \,{\tilde {\hat F}}_{M N} +
( c \, \chi + d)\,{\hat F}_{M N}\nnu\\
&& a\,d - b\,c = 1
\label{30.4}
\ega
where
\bge
{\tilde{\hat F}}_{M N}= {1 \over 2} \sqrt{\det {\hat G}_{M N}}
\,\,\epsilon_{M N R S} {\hat F}^{R S}.
\nnu
\ege
While we consider S-duality transformations in four dimensions,
it is more convenient to work in the Einstein frame rather than
working in the string ( $\sigma$-model ) metric. Therefore it is
to be understood that whenever we discuss S-duality in this paper
we are in the Einstein frame.
It is worthwhile to consider the above $SL(2,{\bf Z})$ transformations
on the equations of motion. We find that the equations of motion are
not invariant under these transformations due to the presence of
the cosmological constant term. We may remark in passing that if
cosmological constant were to have vanishing value the string
equations of motion would have have been invariant under
S-duality.

Next we consider the Euclidean string effective action
   in four-dimension

   \bge
   S_E = \int d^4X^M \sqrt{\det {\hat G}^E_{M N}}\,
   e^{-{\hat\Phi}}\left( R_{\hat G}+{\hat G}^{E \,M N}
\partial_M{\hat\Phi}\partial_N{\hat\Phi} -{1\over4}{\hat F}^{M
N}{{\hat F}}_{M N}
 -2 \Lambda \right)
\label{31s}
   \ege
   where $M = 1,\cdots,4$ and  ${\hat G}^E_{M N}$ is an Euclidean metric.
   As mentioned earlier, the presence of dilaton imposes constraints on
background. In this case the Weyl tensor is anti-self dual. This is as
close as one can get to self-dual Ricci tensor. Strictly speaking one
demands self-duality or anti-self-duality while looking for instanton like
solutions. It is necessary to introduce a self-dual Abelian field
strength in (\ref{31s}) in order to satisfy Einstein as well as matter field
equations.  The background fields are given by

\bga
{\hat ds}^2 &=& { dr^2 \over f(r)^2} + { r^2 \over 4 f(r)^2} ( d\psi +
\cos\theta d\phi)^2 + {r^2 \over 4 f(r)} ( d\theta^2 + \sin^2
\theta d\phi^2) \nnu\\
{\hat A}_M &=& ( \,\,0,\,\,0,\,\, \sqrt{\Lambda\over2}{r^2 \cos\theta
\over 2 f(r)}\,\,, \,\,\sqrt{\Lambda\over 2} { r^2 \over 2 f(r)}
\,\,) \nnu\\
{\hat\Phi} &=& {\rm constant}
   \label{31}
\ega
   where $f(r) = 1 + {\Lambda\over 6} r^2$ . This is the
Fubini-Study metric for $CP^2$ manifold and the solution has the
interpretation of gravitational instantanton \cite{gh}.
The Weyl tensor satisfies the anti-self-duality condition
   \bge
   C_{M N P Q} = -\,^{\ast} C_{M N P Q} = - {\sqrt{\det G^E}\over 2}
\,\epsilon_{M N R S} C^{R S}_{\,\,\,\, P Q}\,.
   \nnu
   \ege
Note, however, Gibbons-Pope instanton solution \cite{gh} was obtained for pure
gravity in presence of cosmological constant, $\Lambda$, whereas action (21)
 contains a gauge field and the dilaton.

The solution (\ref{31}) is a stringy background with vanishing
antisymmetric tensor field strength. All background fields are
independent of two periodic coordinates, $0\le \phi\le 2\pi$ and
$ 0 \le \psi \le 4 \pi$. It is obious that the action
(\ref{31s}) is invariant under $ O(2,3)$ transformations since
the backgrounds are independent of $\phi$ and $\psi$. We can
exploit this symmetry of the action to obtain new classical
backgrounds with nonvanishing antisymmetric tensor field, by
adopting a procedure similar to those discussed above. The first step
is to construct the corresponding moduli matrix $M$ for the
problem at hand. The various moduli fields which are used to construct
M are

\bga
G_{\alpha\beta}&=&\pmatrix{ {r^2\over4f(r)^2}(\cos^2\theta +
f(r)\sin^2\theta) & {r^2\over4f(r)^2}\cos\theta \cr
{r^2\over4f(r)^2}\cos\theta & {r^2\over4f(r)^2}}\nnu\\\nnu\\
A_\alpha&=&(\,\,\, \sqrt{\Lambda\over2}{r^2\cos\theta\over2
f(r)}\,\,\,,\,\,\, \sqrt{\Lambda\over2}
{r^2\over2f(r)}\,\,\,)\nnu\\
B_{\alpha\beta}&=&\,0,
\label{33}
\ega
where indices $\alpha\,,\,\beta$ run over coordinates ( $\phi$, $\psi$ )
in the given order. The next step is to choose an $O(2,3)$ matrix $\Omega$. To
generate inequivalent backgrounds from the one given in
(\ref{31}) we choose $\Omega$ to be a special element of
$O(2)\times O(3)$ , which is a subgroup of $O(2,3)$ ,
given by
\bge
\Omega= \pmatrix{ 1&0&0&0&0\cr 0&0&0&1&0\cr0&0&1&0&0\cr0&1&0&0&0
\cr0&0&0&0&1}\,.
\label{32}
\ege
Now one can perform the transformations in (\ref{17}) using
eqs.(\ref{32}) and (\ref{33}) above. We get the new field
 configuration
 \bga
 {\bar G}_{\alpha\beta}&=& \pmatrix{ {r^2\sin^2\theta \over 4f(r)} & 0 \cr
 0 & { 4 f(r)^2\over r^2(1 + {\Lambda\over4} r^2)^2}}\nnu\\\nnu\\
 {\bar A}_\alpha&=& (\,\,\, 0\,\,\,, \,\,\, -\sqrt{2\Lambda} {f(r)\over(1 +{
 \Lambda\over4} r^2)} \,\,\,)\nnu\\
 {\bar B}_{\alpha\beta}&=& \pmatrix{ 0&\cos\theta \cr -\cos\theta & 0}\nnu\\
{\bar \Phi}(r)&=& {\hat \Phi} - \ln \left( {r^2
(1+{\Lambda\over4}r^2)\over4(1+{\Lambda\over6}
 r^2)^2} \right)\,.
 \label{34}
 \ega
Interestingly, we have got a new field configuration in which
the metric is diagonal. Also, dilaton and antisymmetric tensor
fields are nontrivial.
 It should be noted that both Weyl tensor as well as vector
field strength have lost their respective self-duality
 properties after this transformation.
 Asymptotically ( i.e., as $r\to\infty$ ) the new four-metric has the form
 \bge
{\bar ds}^2 = \omega(r)^2( dr^2 + r^2 d\Theta^2) +
{6\over4\Lambda} ( d\theta^2 + \sin^2\theta \,d\phi^2)
 \label{36}
 \ege
where $\omega(r) = {6\over \Lambda r^2}$ and $\Theta =
{2\Lambda\over 9}\psi$. Let us for the sake of simplicity take
$\,{\hat \Phi}=0\,$ and discuss the behaviour of
$\,{\bar\Phi}\,$ as a function of $r$. Asymptotic value of the
dilaton, ${\bar\Phi}_{asym}=-\ln{9\over4\Lambda}$, depends on
the cosmological constant $\Lambda$ . Notice that for $r\sim0$,
$\,{\bar\Phi}_{r\sim0}\sim -\ln{r^2\over4}$.  We recall that
$\,e^{\bar\Phi}\,$ is the string coupling constant. It is
 interesting to note that string coupling has ${1\over r^2}$ behaviour near
the origin while attains a constant value of $\,4 \Lambda \over 9\,$ at
$ r\to \infty$.
We observe that for large $r$, $\,\,e^{\bar\Phi}\,$ is a constant, if
$\Lambda\to 0$ the string coupling tends to vanishing value. Thus for
this simple model under consideration, there is a connection between
string coupling constant and the cosmological constant.

To summarize, in this work we have studied the solutions
of four-dimensional heterotic string
effective actions in presence of
cosmological constant, $\Lambda$.  The solutions of equations
of motion corresponding to the Minkowskian string effective
action are analogous to ``electrovac'' solution of Freedman and
Gibbons \cite{fg}. This action has an invariance under $O(2,3)$
transformations.
Using this property, we could generate new backgrounds with
nontrivial $B_{M N}$. However, the curvature scalar
corresponding to the new metric is singular while initial geometry
had constant curvature.

The Euclidean string action in four dimensions with $CP^2$
geometry and an Abelian self-dual gauge field strength admits
``gravitational instanton'' solution. This action also possesses
an $O(2,3)$ invariance.  We have generated new backgrounds such
that the metric is diagonal and dilaton as well as
antisymmetric fields  acquire coordinate dependence. Furthermore,
Weyl tensor and $F_{M N}$, corresponding to new  background, are
no longer ( anti ) self-dual .

We note that, due to the presence of the cosmological constant
term in the action the equations of motion are not invariant
under the $SL(2,{\bf Z})$ transformations. This
is an interesting feature that S-duality in not respected in
presence of $\Lambda$ for both the problems we have considered.

\acknowledgements { We are thankful to the anonymous referee for
his constructive comments in improving this manuscript.}

\end{document}